\include{wrapfig}

\documentclass[
    ,final            % use final for the camera ready runs
  ]
  {aipproc}

\layoutstyle{6x9}

\begin{document}

\title{The search for missing baryon resonances}

\author{U. Thoma}{
  address={2. Physikalisches Institut, University of Giessen, \\
Heinrich-Buff-Ring 16, 35392 Giessen, Germany}
}

\begin{abstract} 

Experiments with electromagnetic probes are promising 
to search for baryon resonances that have been predicted 
by quark models but have not yet been observed. 
Data sets from different experiments show interesting
resonance structures possibly due to so far unknow 
states. This might indicate that at least some of the missing baryon
resonances start to show up. 
\end{abstract}

\maketitle

%%%%%%%%%%%%%%%%%%%%%%%%%%%%%%%%%%%%%%%%%%%%
%% MAINMATTER
%%%%%%%%%%%%%%%%%%%%%%%%%%%%%%%%%%%%%%%%%%%%

\subsection{Introduction} 
At medium energies, our present understanding of QCD is still very
limited. Here, in the energy regime of meson and baryon 
resonances the strong coupling constant is large and 
perturbative methods can no longer be applied. 
One of the key issues in this energy regime is to identify
the relevant degrees-of-freedom and the effective forces between them. 
A necessary step towards this aim is undoubtedly a 
precise knowledge of the experimental spectrum of baryon resonances
and of their properties.  
Their comparison with different models may lead to a deeper
understanding of QCD in the non-perturbative regime.   
Quark models are in general amazingly successful in
describing the spectrum of existing states. 
However, constituent quark models usually predict many more resonances 
than have been observed so far. 
Different explanations have been suggested to explain this observation: \\
1) The "missing" states may not exist. 
Nucleon resonances could e.g. have a
quark-diquark structure~\cite{lichtenberg}. This reduces the number of internal 
degrees-of-freedom, and therefore, the number of existing states.
Of course, one might also think of other hidden symmetries. 
At a first glance, this explanation seems to be rather exotic but  
%QCD does
%not give us any reason to believe that the three quarks shouldn't play
%an equal role in the nucleon. 
%
%
it is interesting to notice that the Regge trajectories
for mesons and baryons are parallel. 
The similar dependence of the mass squared on the angular momentum
seems to indicate that also the acting force is similar. This
behavior could be easily understood in terms of a quark-diquark
picture, with a diquark in a baryon replacing the antiquark in the
meson (see also~\cite{klempt_massformula}). \\
2) The "missing" states may not have been
observed so far because of a lack of high quality data in channels
different from $\pi N$. Most available experimental data stem from
$\pi N$ scattering experiments. If the missing states decouple from $\pi N$
they would not have been observed so far. This conjecture seems reasonable
following quark model predictions~\cite{capstick9394}. 
Many of these unobserved states are expected to couple significantly
to  channels like $N\eta$, $N\eta^{\prime}$, $N\omega$, $\Delta \pi$, $N\rho$ or
$K\Lambda$ and also to $\gamma
p$~\cite{capstick9394,capstick92}. Therefore photoproduction 
experiments investigating these channels have a great discovery
potential if these resonances really exist. \\
Experiments with electromagnetic probes are not only interesting to search for 
unknown states  but also to determine the properties of resonances like
photo-couplings and partial widths. These provide additional
information which can be compared to model predictions.  
The properties of a resonance are also of big importance for an
interpretation of its nature. One immediate debate in the light of the
possible observation of a pentaquark is e.g. whether the $\rm P_{11}$(1710) and
the $\rm P_{11}$(1440) might be pentaquarks rather than 3-quark states. A good
understanding of their production and decay properties may help to elucidate
their nature. 
In the following, different final states, where interesting resonance
structures have been observed, will be disucssed. \\[-3ex]  
\subsection{The \boldmath$\gamma p \to p \eta$\unboldmath-channel}
\begin{figure}[h!]
\begin{tabular}{rl}
  {\includegraphics[width=.425\textwidth,angle=0]{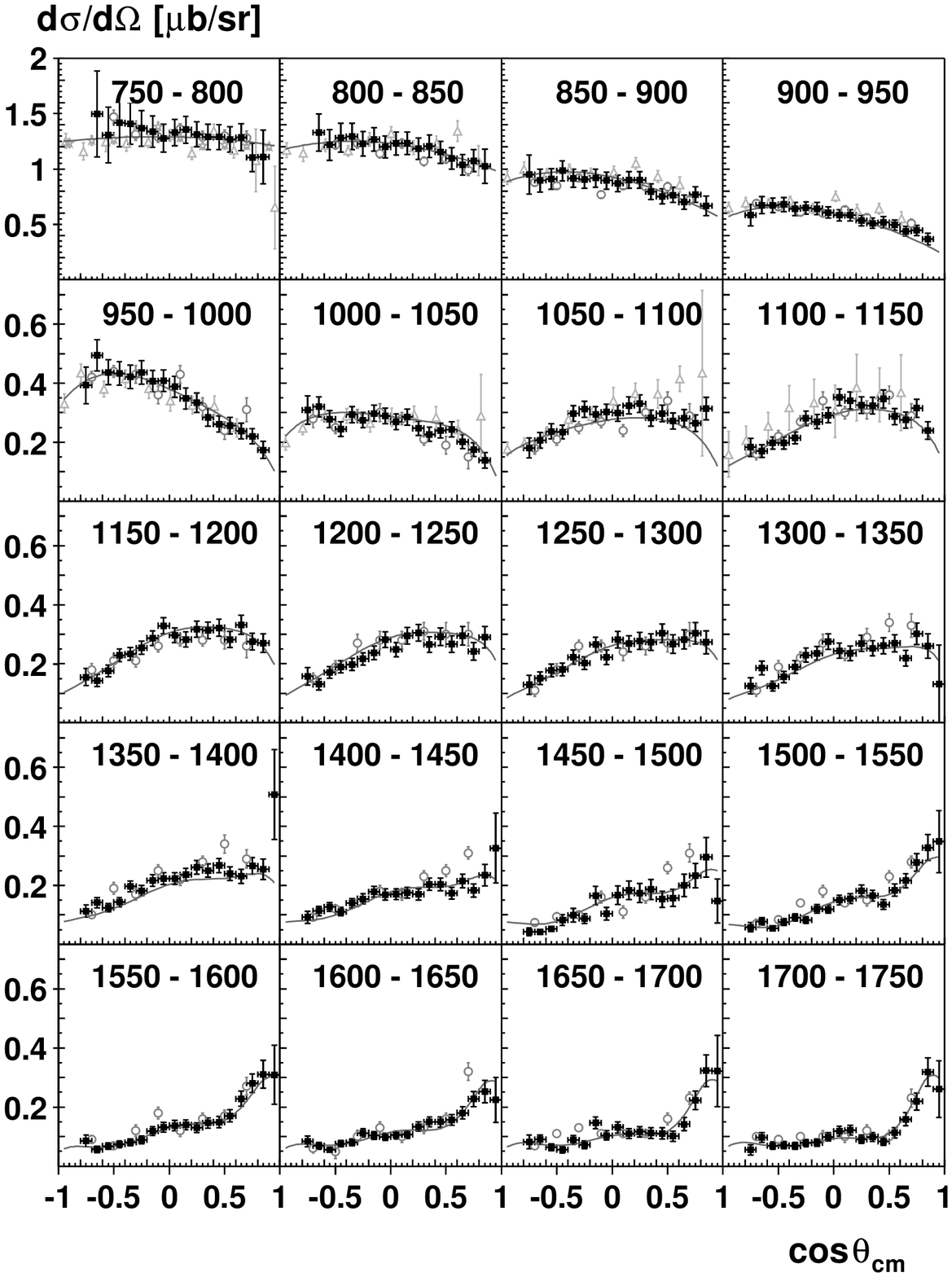}} &   \\[-40ex] 
& {\includegraphics[width=.425\textwidth,angle=0]{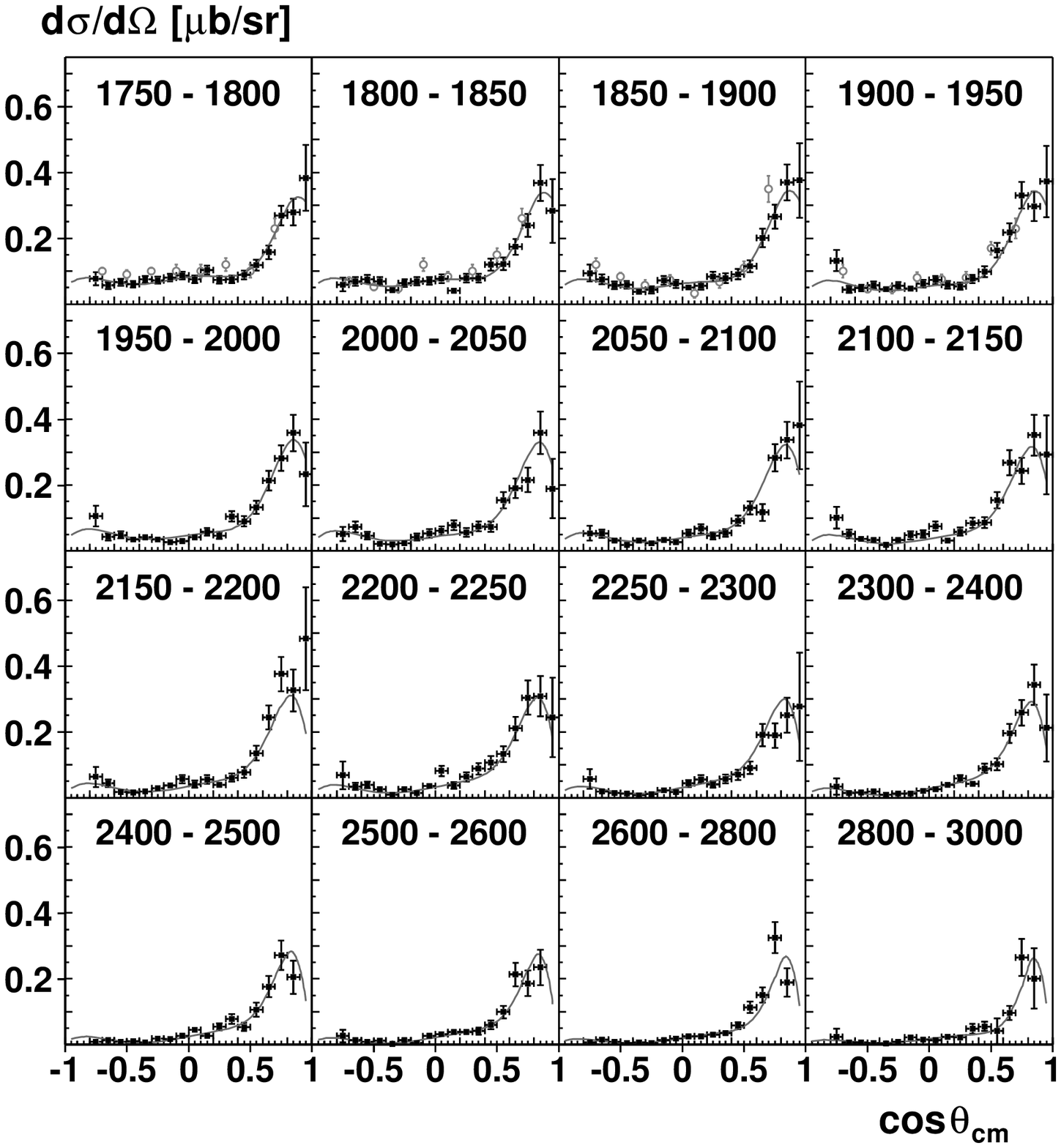}}  \\[-2ex] 
{\includegraphics[width=.47\textwidth]{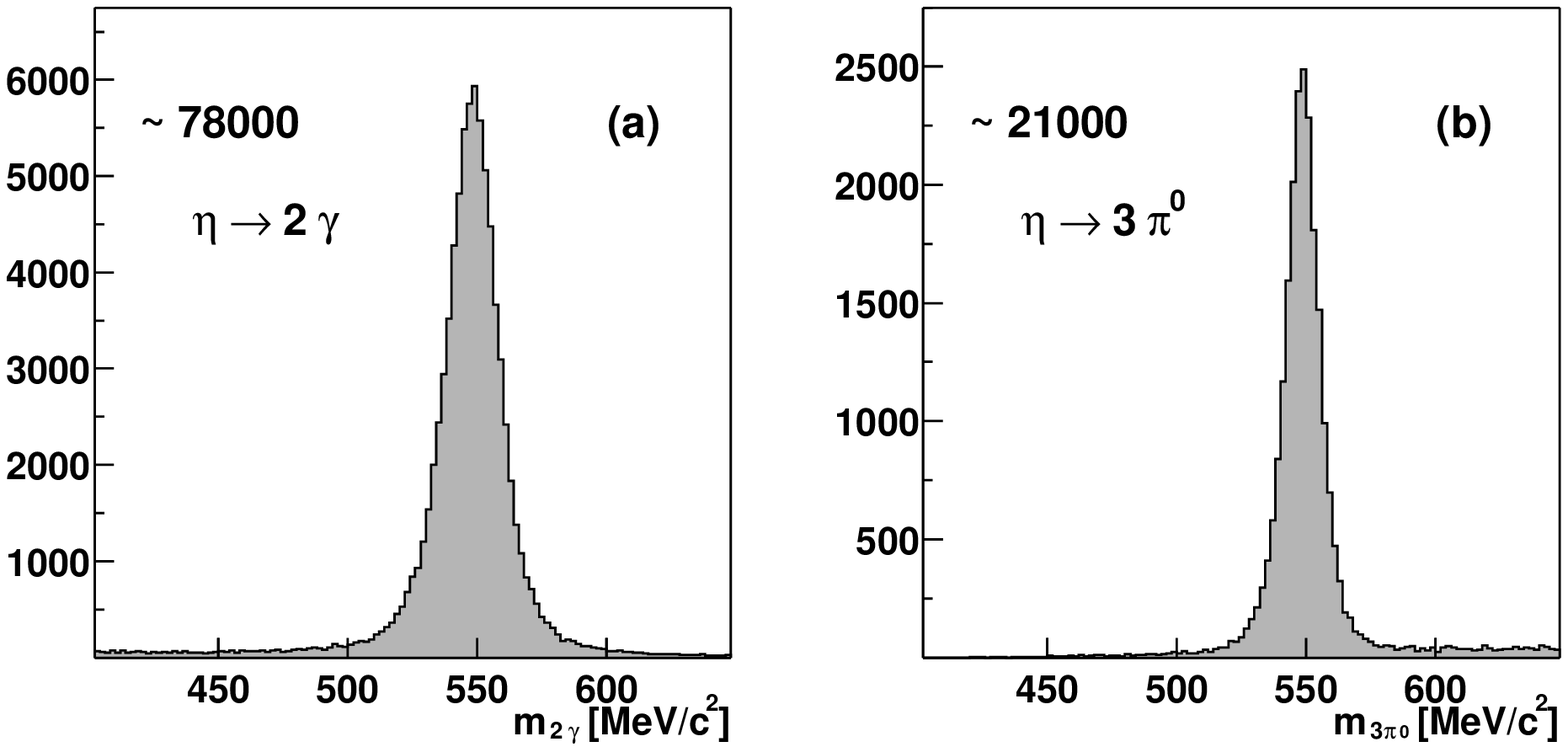}} 
&  \hspace*{+0.1cm} {\includegraphics[width=.42\textwidth]{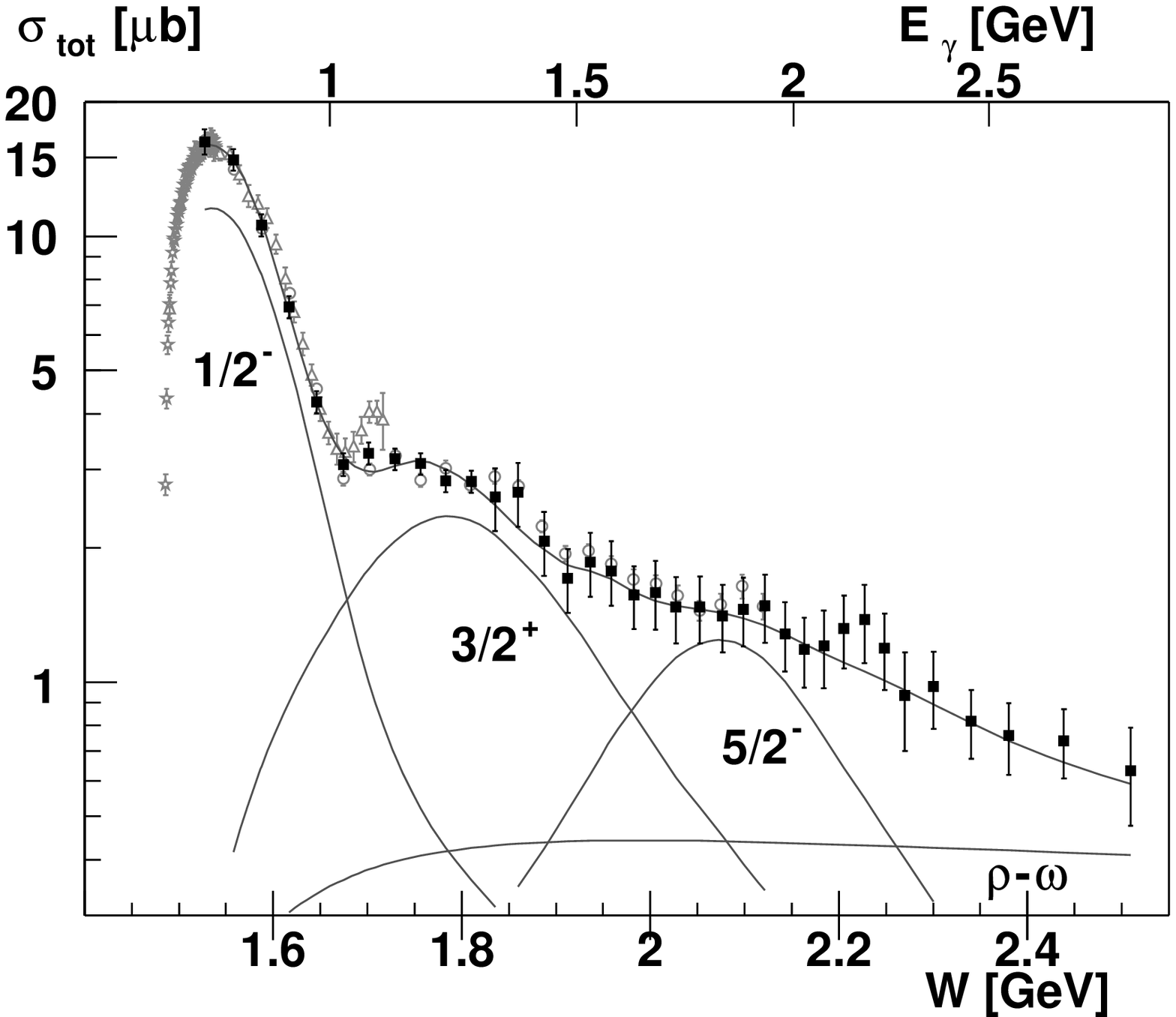}} \\ %[+7ex]
\end{tabular}
\caption{Upper plots: Differential cross sections for $\gamma\,\rm{p}
\rightarrow \rm{p}\,\eta$, for $E_\gamma = 750$\,MeV to
3000\,MeV: CB-ELSA(black squares)~\protect\cite{eta_pap}, TAPS~\protect\cite{Krusche:nv},
GRAAL~\protect\cite{Renard:2000iv} and CLAS~\protect\cite{Dugger:ft} data (in
light gray). The solid line represents the result of our fit.
Lower left: Invariant $\gamma\gamma$ and $3\pi^0$ invariant mass. 
Lower right: Total cross section (logarithmic scale) for the reaction
$\gamma\,\rm{p}\rightarrow\rm{p}\,\eta$. %The contributions of the two
%S$_{11}$ resonances, of the $\rm N(1720)P_{13}$, of the $\rm
%N(2070)D_{15}$, and of the background amplitudes (mainly $\rho-\omega$
%exchange) are shown as well. 
For further details see~\protect{\cite{eta_pap}}.
 }
\label{fig_eta}
\end{figure}

Recently new data on $\eta$-photoproduction has been taken by the
CB-ELSA experiment in Bonn % extending the covered photon energy range up
%to $E_{\gamma}$=3.0$\,$GeV
\cite{eta_pap}. 
Due to its electromagnetic calorimeter consisting of 1380 CsI(Tl)
crystals covering 98$\%$ of the 4$\pi$ solid angle, the CB-ELSA
detector is very well suited to measure
photons.  The $\eta$ is observed either in its $\gamma\gamma$- or 3$\pi^0$-
decay. The two or six photons are
detected in the calorimeter and the proton is 
identified in a 3-layer scintillating fiber detector. 
The invariant masses show a clear $\eta$ signal over
an almost negligible background (Fig.~\ref{fig_eta}). 
The differential as well as the total cross section is shown in
Fig.~\ref{fig_eta} in comparison to the TAPS~\cite{Krusche:nv},
GRAAL~\cite{Renard:2000iv} and CLAS~\cite{Dugger:ft} data. The new CB-ELSA data
extends the covered angular and energy range significantly compared to previous
measurements. The total cross section was obtained by integrating the
differential cross sections. The extrapolation to forward and 
backward angles uses the result of the partial wave analysis (PWA) discussed below.
%The solid line represents the result of the  partial wave analysis.
The PWA is necessary to extract the contributing resonances 
from the data. Its result is shown as   solid line  in
Fig.~\ref{fig_eta}. 
In the fit the following data sets were included in addition to the CB-ELSA
data on $\rm \gamma p \rightarrow p\eta $: The CB-ELSA data on $\rm \gamma p
\rightarrow 
p\pi^0 $ \cite{Bartholomy:04}, the TAPS data on $\rm \gamma p
\rightarrow p\eta $ \cite{Krusche:nv}, the beam asymmetries
$\rm\Sigma(\gamma p \rightarrow p\pi^0)$ and $\rm\Sigma(\gamma p
 \rightarrow p\eta)$ from GRAAL \cite{GRAAL2}, 
and $\rm\Sigma(\gamma p \rightarrow p\pi^0)$ and 
$\rm \gamma p \rightarrow n\pi^+ $ from SAID. 
Apart from known resonances a new state was found, an $\rm
D_{15}(2070)$ 
with a mass of ($2068\pm 22$)\,MeV and a width of ($295\pm 40$)\,MeV. Its
rather strong contribution to the data set is also shown in
Fig.\ref{fig_eta}. In addition an indication for a possible new $\rm
P_{13}(2200)$ was found. 
No evidence was found for a third S$_{11}$ for which claims have been
reported at masses of 1780\,MeV~\cite{Saghai:2003ch} and
1846\,MeV~\cite{Chen:2002mn}. \\[-3ex]  
\subsection{The \boldmath$\gamma p \to K^+ \Lambda$\unboldmath-channel }
\begin{figure}[t!]
\begin{tabular}{cc}
\vspace*{-0.0cm}
{\includegraphics[height=.255\textwidth]{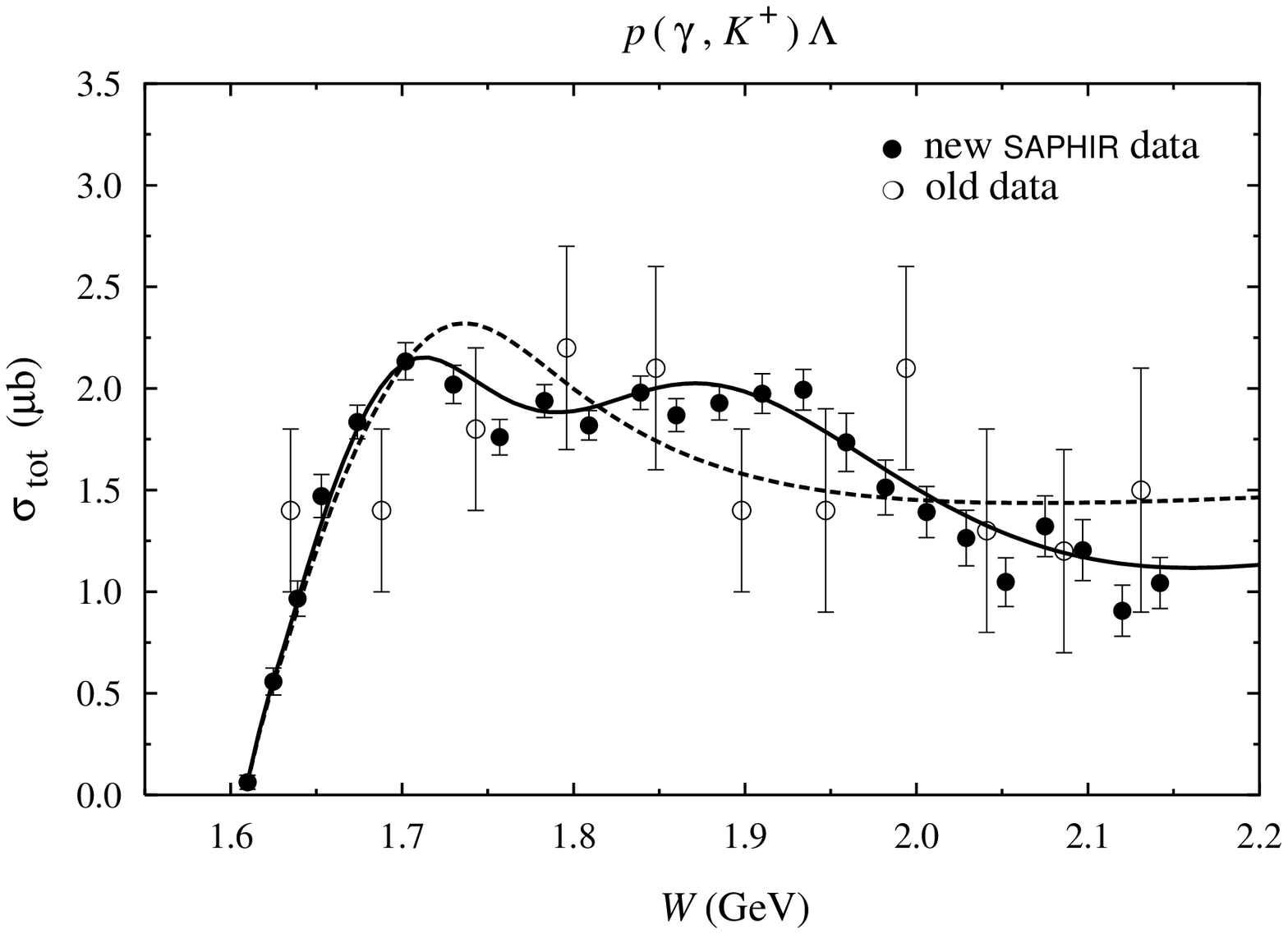}}   & \\[-20ex]
& \hspace*{-.cm}{\includegraphics[height=.57\textwidth]{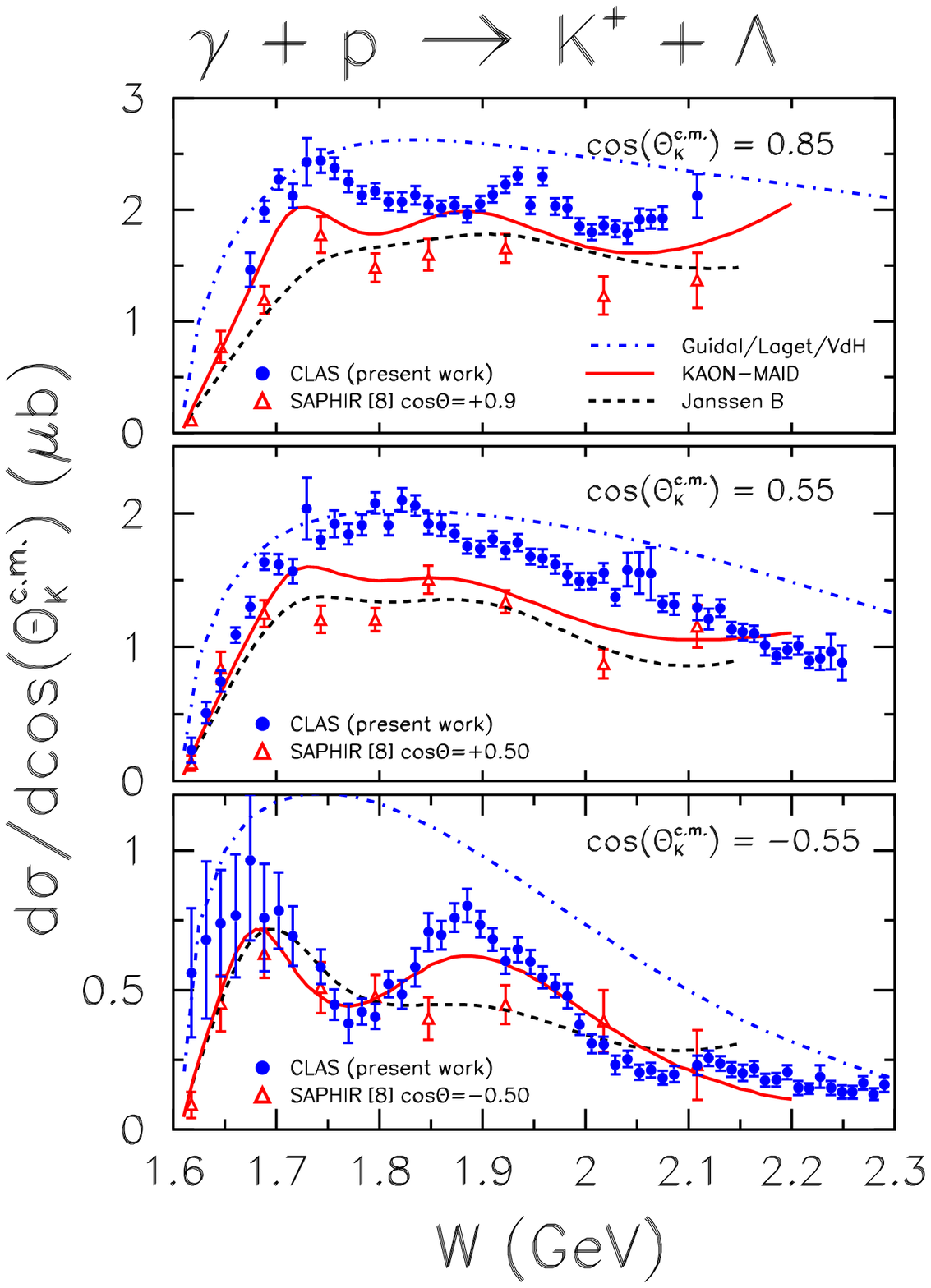}}   \\[-28ex]
\hspace*{-0.2cm}{\includegraphics[height=.235\textheight]{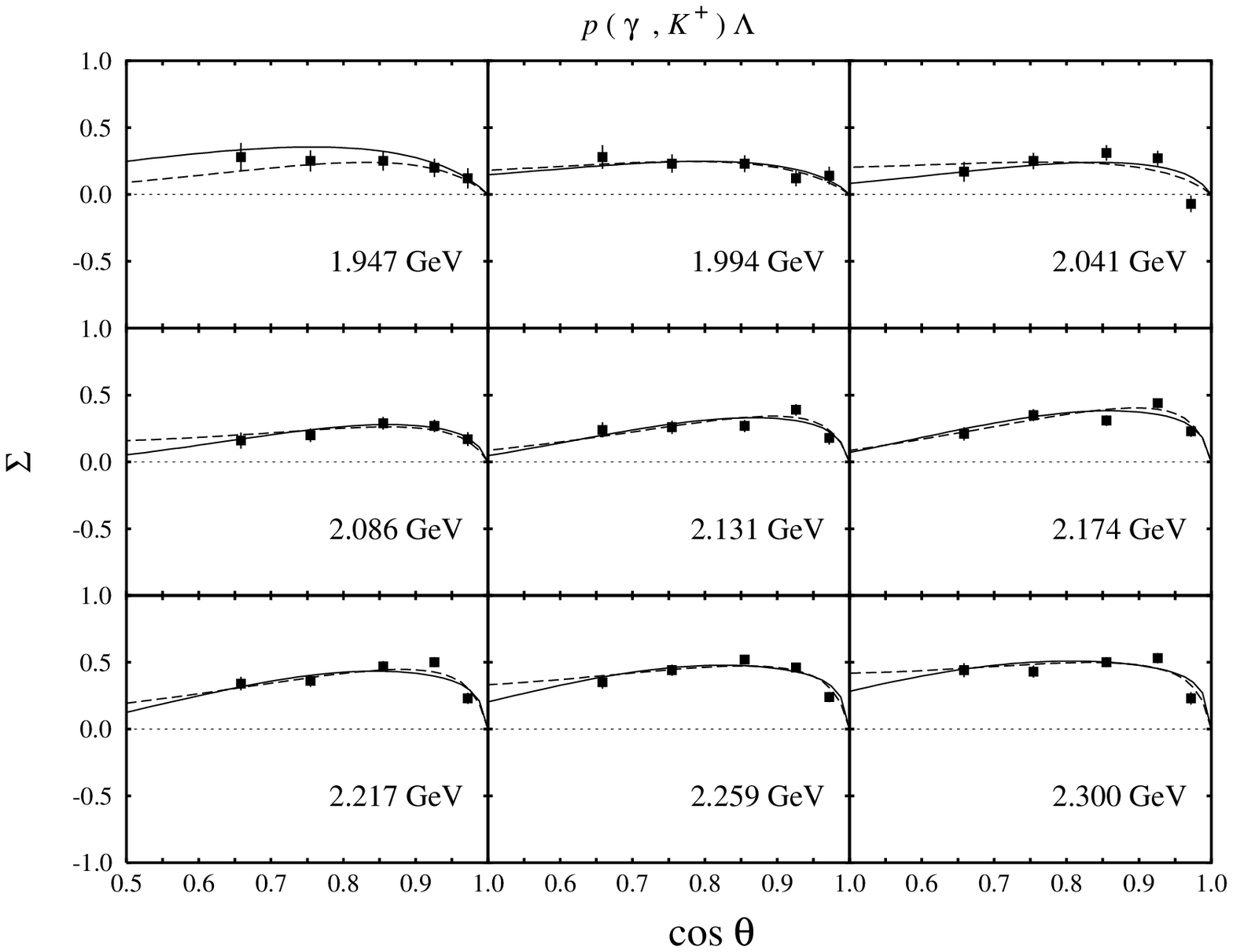}}
& \\
\end{tabular}
\caption{
  Left, upper plot: Total cross section for $\gamma p \to K^+ \Lambda$ from SAPHIR
  \protect{\cite{saphir_old}}. Description of the data within the model~\protect{\cite{mart_bennhold}} with and without including
  a new resonance   around 1900~MeV. 
  Right: Energy dependence of the $\gamma p \to K^+ \Lambda$-cross
  sections for different $K^+$-angles   (CLAS-data\protect{~\cite{klambda_clas}}). 
  Left, lower plot: Beam asymmetry for $\gamma p \to K^+ \Lambda$ measured by
  LEPS~\protect{\cite{klambda_leps}}. }
\vspace*{-0.3cm} 
\label{klambda}
\end{figure}

Another interesting channel is the $K^+\Lambda$ channel, where a 
structure around 1900~MeV was first observed in the SAPHIR 
data~\cite{saphir_old}. The total cross section  shows two bumps, at about 1700~MeV, and 1900~MeV (Fig.\ref{klambda} upper left). 
%(Fig.\ref{klambda}, upper left). 
%\begin{figure}[t!]
%\begin{tabular}{c}
%  {\includegraphics[height=.3\textwidth]{saphir_d13.eps}} 
%\end{tabular}
%\caption{Total cross section for $\gamma p \to K^+ \Lambda$ from SAPHIR
%  \protect{\cite{saphir_old}}. Description of the data within the model of
%  Mart and Bennhold~\protect{\cite{mart_bennhold}} with and without including
%  a new resonance   around 1900~MeV. }
%\label{klambda_saphir}
%\end{figure}
Describing the data within different models it was
found that the first peak is mainly due to the $\rm S_{11}(1650)$, 
$\rm P_{11}(1710)$, and $\rm P_{13}(1720)$. Within a tree
level isobar model based on an effective lagrangian approach Mart and Bennhold
found that a new resonance is needed to describe the second bump in the cross
section~\cite{mart_bennhold}. This resonance was identified with a
$\rm D_{13}$(1895). Fig.\ref{klambda}, upper left shows their best description of
the data with and without this new state. 
Even though this picture looks quite convincing the existence of the
state is controversially discussed. 
Saghai was e.g. able to describe the data within a chiral quark
model without any new resonance~\cite{klambda_saghai}. In his model the enhancement is 
explained by hyperon exchange in the u-channel. 
Hyperon exchanges are also included in the model of Jannsen et
al.~\cite{klambda_jannsen} but still an additional state around 1900
MeV was needed. A similar improvement of the fit was obtained
by resonances of different quantum numbers.  
In the Giessen coupled-channel model a negligible  $K
\Lambda$ coupling was found for a $\rm D_{13}$ state
which was introduced around 1950 MeV~\cite{klambda_giessen}.
Recently new high statistics data on this final state became available. 
SAPHIR~\cite{klambda_saphir} and CLAS~\cite{klambda_clas} did provide new data on cross
sections and on the $\Lambda$ recoil polarization and  
LEPS~\cite{klambda_leps} on the 
beam-asymmetry. % obtained by using linearly polarised photons.  
The differential cross sections as a function of $\sqrt{s}$ for
different K-angles are shown in Fig.\ref{klambda}, right for the CLAS data. 
Again a structure around 1900\,MeV is observed. It varies in width and
position with the $K^+$-angle. This suggests an interference phenomenom between
several resonant states, rather than a single
resonance. This behaviour is not yet reproduced by the models shown, but the
model parameters have not yet been adjusted to the new data. 
Some first prelimary results on an interpretation of the new data 
have been shown by Mart and Bennhold~\cite{mart_bennhold_conf}. 
Fitting the new SAPHIR data together with the beam asymmetry data from
LEPS (Fig.\ref{klambda}, lower left) they find that more than one resonance is needed to
describe the mass region around 1900~MeV. This work is still in progress, so 
no definite statement on the existence of new resonances in this data could be
made yet. \\[-3ex]  
\subsection{The \boldmath$\gamma p \to p \pi^0\pi^0$\unboldmath-channel }
\begin{figure}[t!]
  {\includegraphics[height=.2\textheight]{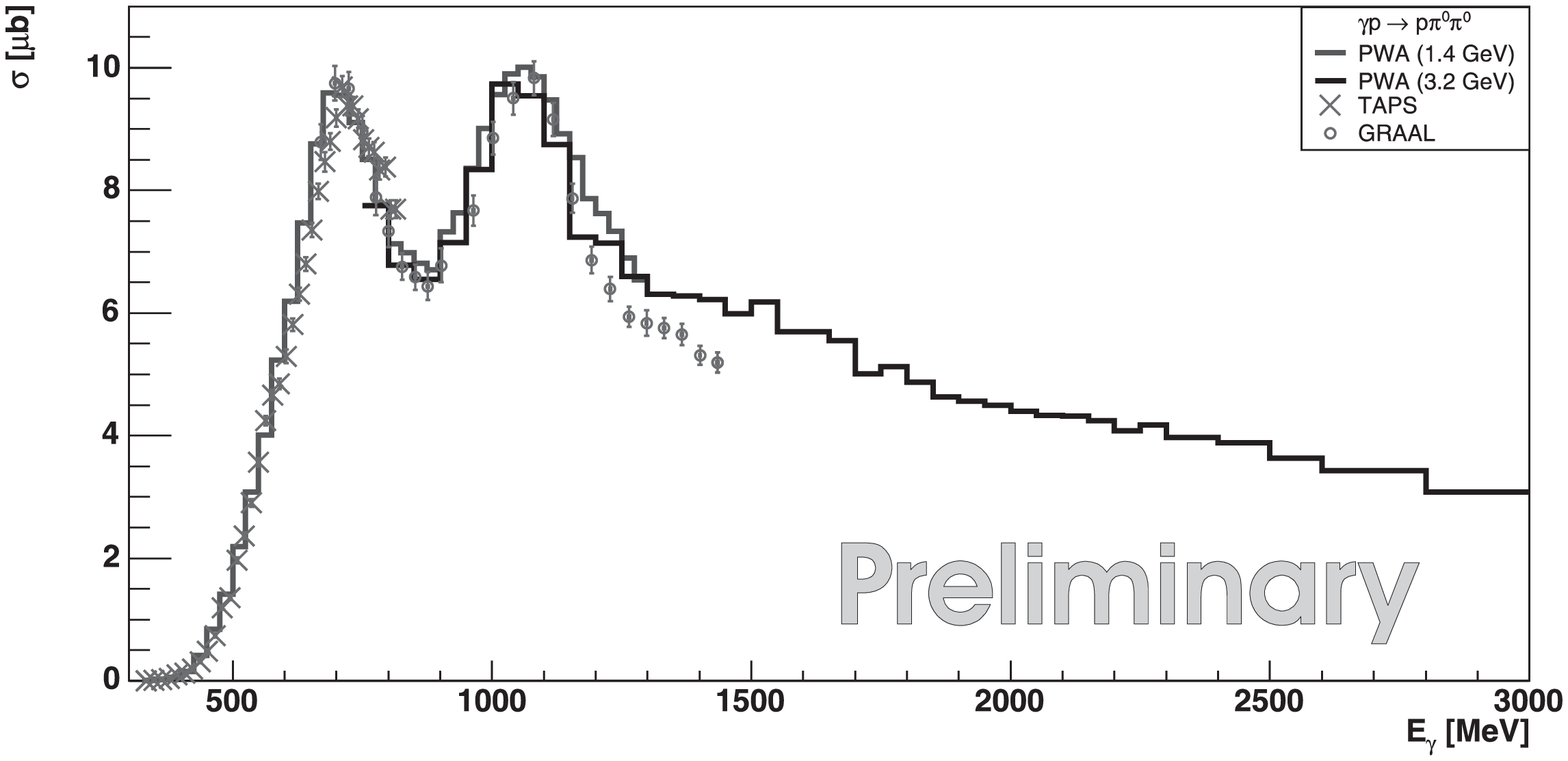}}
  {\includegraphics[height=.2\textheight]{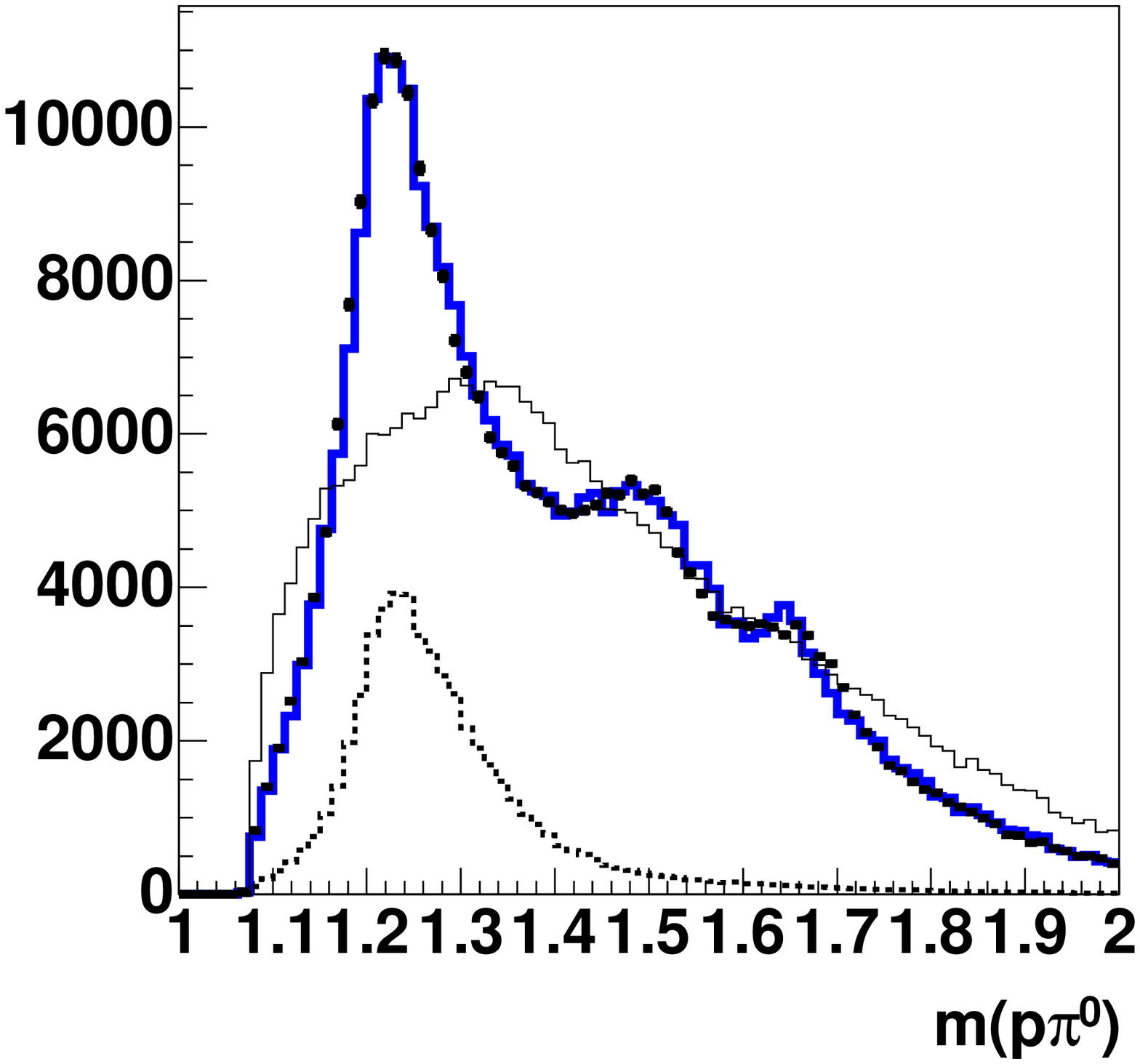}}
\caption{Left: Total cross section as obtained by integrating the result of
the partial wave analysis over phase space (solid line), in comparison
to the preliminary TAPS\protect\cite{kottu} and GRAAL\protect\cite{graal}
data. Right: $p\pi^0$ invariant mass for $E_{\gamma}$=0.8-3.0~GeV  
in comparison to the result of the PWA. 
The plots shows the experimental data (points with error bars), the
result of the PWA  (solid gray curve), the contribution of the $\rm
  D_{13}$(1520) (dashed black curve) and the phase
  space distribution (thin black line), preliminary.}
\label{xsec_high}
\end{figure}

The $\gamma p \to p \pi^0\pi^0$ cross section was measured by TAPS~\cite{taps}
in the low energy range and by GRAAL~\cite{graal} up to an incoming photon
energy of about 1500~MeV; two peak-like structures are observed~\cite{taps,graal}. 
The data has been interpreted within the Laget-~\cite{laget_graal} and Valencia
model~\cite{oset_graal_pap}, 
resulting in very different interpretations. In the Valencia-model, which is
limited to the low 
energy region, the $\rm D_{13}$(1520) decaying into $\Delta (1232)\pi$
dominates the lower energy peak, while
in the Laget-model the  $\rm P_{11}$(1440) decaying into $\sigma p$ is clearly
the dominant contribution. Even though both models lead to a
reasonable description of the total cross section their interpretation of the
data is rather different, in fact they are  contradicting each other. 
Recently data on $\rm \gamma p\to p \pi^0\pi^0$ has also been taken by the
CB-ELSA experiment in Bonn extending the covered energy range up
the $E_{\gamma}$=3.0$\,$GeV~\cite{cbelsa_ppi0pi0}. 
To extract the contributing resonances, their quantum numbers and their
properties from the data, a PWA has been done. 
The formalism used is summarized in~\cite{formalism}. The fit uses Breit-Wigner
resonances and includes  $s$- and $t$-channel amplitudes. 
An unbinned maximum-likelihood fit was performed which has
the big advantage of being event-based; it takes all the 
correlations between the five independent variables correctly into account. 
The fits include the preliminary TAPS data~\cite{kottu} in the low energy region
in addition to the CB-ELSA data. 
%The latter was taken using two different energy 
%settings ($E_{e^-}$=1.4~GeV, 3.2~GeV).  
%
Resonances with different quantum numbers were
introduced in various combinations allowing, so far, for the following decay
modes: $\Delta(1232)\pi$, $\rm N(\pi\pi)_s$, $\rm P_{11}(1440)\pi$, $\rm
D_{13}(1520)\pi$ and $\rm X(1660)\pi$.
For a good description of the data resonances like e.g. the $\rm
P_{11}(1440)$, the $\rm D_{13}(1520)$, the $\rm D_{13}/D_{33}(1700)$, the $\rm
P_{13}(1720)$, the $\rm F_{15}$(1680) as well as several additional states are
needed. 
One preliminary result of the PWA is a dominant
contribution of the  $\rm D_{13}(1520)\to \Delta \pi$
amplitude in the energy range, where the first peak in the cross
section occurs. 
%This is in contradiction to the interpretation given within the Laget model.
%where the $ P_{11}(1440)\to p\sigma$-amplitude dominates in this energy range. 
%%
%
Fig.~\ref{xsec_high} shows the total cross section obtained by fitting
the CB-ELSA and the TAPS data and by integrating the result of
the combined fit over phase space.   \\
In the CB-ELSA data baryon resonances not only decaying into $\Delta \pi$ but
also via $\rm D_{13}(1520)\pi$ and $X(1660)\pi$ are observed for the first time. 
The enhancements at the corresponding $p\pi$ invariant masses
are clearly visible in Fig.~\ref{xsec_high}, right. The observation of baryon cascades is
also interesting with respect to 
the search for states which might not couple to $\pi N$ and $\gamma p$; they
still could be produced in such baryon cascades. \\[-3ex]  
\subsection{The \boldmath$e p \to e^{\prime} p \pi^+\pi^-$\unboldmath -channel }
\begin{figure}[t]
  {\includegraphics[height=.22\textheight,width=.33\textheight]{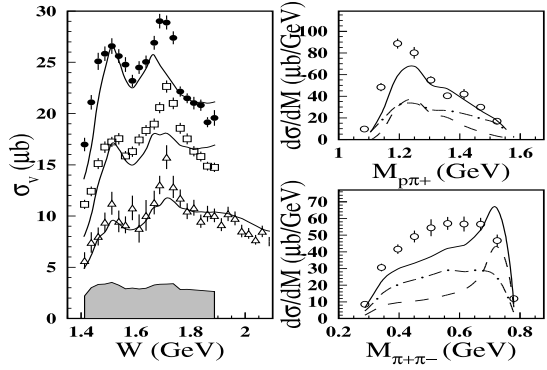}}
  \caption{Left: Total cross section for $\gamma^* p \to p
    \pi^+\pi^-$ for 
  $Q^2$=0.5-0.8 (full points), 0.8-1.1 (open squares), 1.1-1-5 (GeV/c)$^2$
    (open triangles)~\protect\cite{ripani}. Right: Differential cross
    sections for W=1.7-1.75 GeV, $Q^2$= 0.8-1.1 (GeV/c)$^2$. 
    The line corresponds to the fit which includes the known
    information on the resonances described in the text. For further
    details see~\protect\cite{ripani}.       }
\label{clas1}
\end{figure}

Recently,  $2\pi$-electroproduction was investigated by CLAS. % (Jefferson Lab.).
The total cross section for different bins in 
momentum transfer $Q^2$ is shown in Fig~\ref{clas1}. 
The cross section changes with $Q^2$ as one would expect since the
helicity couplings of the resonances depend on $Q^2$. 
The data was investigated within an isobar model. The fit takes the
$\Delta\pi$ and the $\rm N\rho$ subchannels into account, in addition
non-resonant contributions are allowed for. 
The fit included 12 known baryon resonances together with the non-resonant
background amplitudes.  If all known information on the resonances is included
in the fit the mass region 1700~MeV is rather badly described while the fit agrees 
fairly well with the data at low W (Fig.\ref{clas1}). 
At the same time, the fit clearly overshoots the data in the $\rho$ region of
the $\pi^+\pi^-$ invariant mass in the W-bin around 1700~MeV.  
This behavior can be traced back to the $\rm P_{13}$(1720) 
which has, following the PDG, a rather large $\rm N\rho$-decay
width of 70 to 85$\%$. 
A better description of the data is reached by either changing the decay properties
of the known $\rm P_{13}$ completely or by keeping the PDG-state and adding a new $3/2^+$
state of slightly smaller width and  a stronger $\Delta\pi$ decay mode. For further
details see~\protect\cite{ripani}.   
A recent combined analysis of this data together with the CLAS
$p\pi^+\pi^-$-photoproduction data seems to indicate that two state are needed to
describe these two data sets consistently~\cite{victor}. But this analysis is
still in progress. \\[-3ex]  
\subsection{Summary}
Indications for new resonances are found in several finals states.  
A $\rm D_{15}$(2070) and possible indications for an 
$\rm P_{13}$(2200)-state were 
found in a combined PWA of the new CB-ELSA $\eta$-photoproduction
together with other data sets. 
In the old $\rm K^+\Lambda$-SAPHIR data possible indications for a new
state around 1900~MeV were observed, while the new higher statistics 
data on this final state from SAPHIR, CLAS and LEPS seems to indicate that
even more than one resonance might contribute to the mass region around 1900~MeV.  
2$\pi^0$-photoproduction data has been taken by TAPS, GRAAL and CB-ELSA. 
The total cross section shows two clear peaks around $E_{\gamma}$=700~MeV and
1100~MeV. The interpretation of the lower energy peak has been controversially
discussed as either being mainly due to the $\rm P_{11}(1440)\to p
\sigma$, or due to the $\rm D_{13}(1520) \to \Delta \pi$ amplitude. A preliminary
combined PWA of the CB-ELSA and TAPS  
data indicates a dominant contribution of the latter amplitude in this energy
range.  At higher energies, in addition to the $\Delta\pi$ decay of
baryon resonances, their decay via higher mass resonances like e.g. the
$\rm D_{13}$(1520) are observed for the first time. 
This observation opens up a new opportunity to search for baryon resonances
which may decouple from $\rm N\pi$ and $\rm \gamma N$; they still might be
produced in baryon cascades.  
The possible existence of an additional $\rm P_{13}$ state  in
the CLAS $\pi^+\pi^-$-electroproduction data around 1700~MeV also has been
discussed recently.  \\
So the question is, whether the missing resonances are finally
appearing. The $\rm D_{15}$(2070) would nicely fit to one of the missing states
and the same would also be true for the $\rm P_{13}$(2200) and the
$\rm D_{13}$(1895) state, while a 
second $\rm P_{13}$ state in the mass range around 1700~MeV is not expected by the
quark model.  \\
Before the above question can be answered, a better
understanding of the spectrum is  obviously needed.  
%so more work needs to be done. 
%
One very important step towards this aim is the measurement of
polarization observables. They provide additional constraints for models or PWA
used to extract resonance information from the data. This increases
the sensitivity on smaller contributions and helps to distinguish
between ambiguous PWA solutions. 
Such polarization data has been taken recently by various experiments e.g. by 
GRAAL, at MAMI, by CLAS, by LEPS, and by CB-ELSA/TAPS. Additional new
measurements will also take place, like e.g. the double polarization experiments
planned at ELSA.  

\subsection{Acknowledgments}
The author acknowledges an Emmy Noether grant from the Deutsche
Forschungsgemeinschaft. 
%\end{theacknowledgments}
%
%

\end{document}